\documentclass[pra, preprint, showpacs, amsmath]{revtex4}
\usepackage{graphicx}
\usepackage{amsmath, amsthm, amssymb}
\newtheorem{theorem}{Theorem}
\newtheorem{corollary}{Corollary}
\newtheorem{lemma}{Lemma}

\newcommand{\beqn}{\begin{equation}}
\newcommand{\eeqn}{\end{equation}}

\newcommand{\bra}{\left < }
\newcommand{\ket}{\right > }

\newcommand{\slim}{\text{s-}\negthinspace\lim}

\begin{document}

\title{Rigorous conditions for the existence of bound states at the threshold in
the two-particle case.\footnote{Dedicated to Walter Greiner on his
seventieth anniversary.}}


\author{Dmitry K. Gridnev}
\email[Electronic address:] {gridnev@physik.uni-kassel.de}
\author{Martin E. Garcia}
\affiliation{Theoretische Physik, Universit\"at Kassel, Heinrich-Plett-Str. 40,
D--34132 Kassel, Germany}
\begin{abstract}
 In the framework of non-relativistic quantum mechanics and with the help of the Greens functions formalism we study the behavior of weakly
bound states as they approach the continuum threshold. Through
estimating the Green's function for positive potentials we derive
rigorously the upper bound on the wave function, which helps to
control its falloff. In particular, we prove that for potentials
whose repulsive part decays slower than $1/r^{2}$ the bound states
approaching the threshold do not spread and eventually become bound
states at the threshold. This means that such systems never reach
supersizes, which would extend far beyond the effective range of
attraction. The method presented here is applicable in the
many--body case.
\end{abstract}


\maketitle

\section{Introduction}\label{sec-intro}

In many problems of quantum mechanics it is important to know what
happens to the wave function of a system as the bound state
approaches the dissociation (decay) threshold. In particular, how
does the size of the system in the ground state change as the system
becomes loosely bound. Among multiple examples of loosely bound
systems in physics one could mention negative atomic and molecular
ions \cite{hogreve}, Efimov states \cite{efimov} and halo nuclei
\cite{zhukov}, \cite{fedorov}.


For a bound state of the system as it approaches the threshold there
could be two possibilities. The first one is that the probability
distribution given by this bound state spreads, meaning that the
probability to find all particles together in the fixed bounded
region of space goes to zero (the size of the system goes to
infinity).
One observes such spreading in helium dimer \cite{dimer} or in halo
nuclei like $^6$He or $^{11}$Li, which are so loosely bound that two
neutrons are about to leave the system and form dilute nuclear
matter around the core nucleus ($^4$He and $^9$Li respectively)
\cite{zhukov}. The second possibility is that the bound state does
not spread and in this case it eventually becomes a \emph{bound
state at the threshold} (the size of the system remains finite).
This phenomenon is called the eigenvalue absorption. This is the
case for doubly negative ions and proton halos \cite{hogreve},
\cite{fedorov}.

Recall, that for two particles interacting through spherically
symmetric potentials with finite range $\it S$--states always
spread, while all states with non-zero angular momentum become bound
\cite{klaus}. Incidentally, it is natural to conjecture that the
ground state of a multi-particle system with pair interactions of
finite range ($V_{ij} (x) = 0$ for $|x| \geq R$) cannot be bound at
the threshold given that the particles are either bosons or
distinguishable. For fermions with short-range interactions it is
hard to say from general principles whether the ground state would
spread or not. The physical approach in this case is to use some
kind of shell model and to figure out if there is a centrifugal
barrier, which prevents the wave function from spreading.

On the other hand, there are potentials, for which bound states do
not spread at all and when approaching the continuum they give rise
to bound states exactly at the threshold \cite{amer}, \cite{gest}.
In particular, the physically important case of repulsive Coulomb
tail case belongs to this type (see the discussion in
\cite{newton}). Let us illustrate this situation by a simple
example. Consider the square well potential plus a repulsive Coulomb
tail, as in Fig.~1 (left), and imagine the ground state in this
potential as it approaches the threshold. The probability
distribution for this state would remain a confined wave packet
regardless of how small is the binding energy. For zero binding
energy there would be a bound state, which would have a falloff of
the type $\sim \exp(-\sqrt{r})$. On the contrary, if we cut off the
positive tail at some arbitrary distance $R_c$, the ground state
approaching the threshold would eventually spread when the binding
energy is sufficiently small, {\em i.e.} the probability to find the
particle in some bounded region of space goes to zero with the
binding energy. The state ``tunnels'' through the barrier. Note that
this change in the behavior  does not depend on the value of $R_c$,
which could be as large as pleased, so this effect is solely due to
the repulsive Coulomb tail. This requires a special care in
numerical calculations of loosely bound systems, because often in
the calculations the potential becomes effectively cut off.

A rigorous proof of the eigenvalue absorption in the case of a
general short-range potential plus a repulsive Coulomb tail was
given in \cite{gest}. Unfortunately, the approach presented in
\cite{gest}, based on the Green's function expansion, is aimed
specifically at the Coulomb long-range part and does not allow
generalizations, for example, to potentials having long-range parts
of the form $r^{-1} + A r^{-2}$, which may arise in multipole
expansions. Here, we shall consider a more general two-body case.
The analysis in \cite{amer} illuminates the possibilities for
various radially dependent long-range parts. However, the arguments
given there are not rigorous. Finally, we should mention, that the
phenomenon of the ground state absorption was proved rigorously in
\cite{ostenhof} for a three-body system with pure Coulomb
interactions and the infinitely massive core. This makes one
conjecture that at the point of critical charge negative ions have
bound states at the threshold, see the discussion in \cite{hogreve}.
In a forthcoming article we shall give the rigorous proof of this
conjecture for distinguishable particles and bosons \cite{later},
see also the discussion concerning many-body systems in the last
section.

The paper is organized as follows. In Sec.~\ref{sec2} we set the
criterium for the eigenvalue absorption. In Sec.~\ref{sec3} we
derive useful upper bounds for the Green's function. These bounds
can also be used to control numerical solutions. In Sec.~\ref{mres}
we prove our main result saying that potentials decaying slower as
$1/r^2$ give rise to bound states at the threshold. The last section
presents conclusions and a short discussion concerning many-body
systems. Finally, the Appendix contains technical details necessary
for the proof in Sec.~\ref{sec2}.

 \begin{figure}[tb]
\includegraphics[width=6cm]{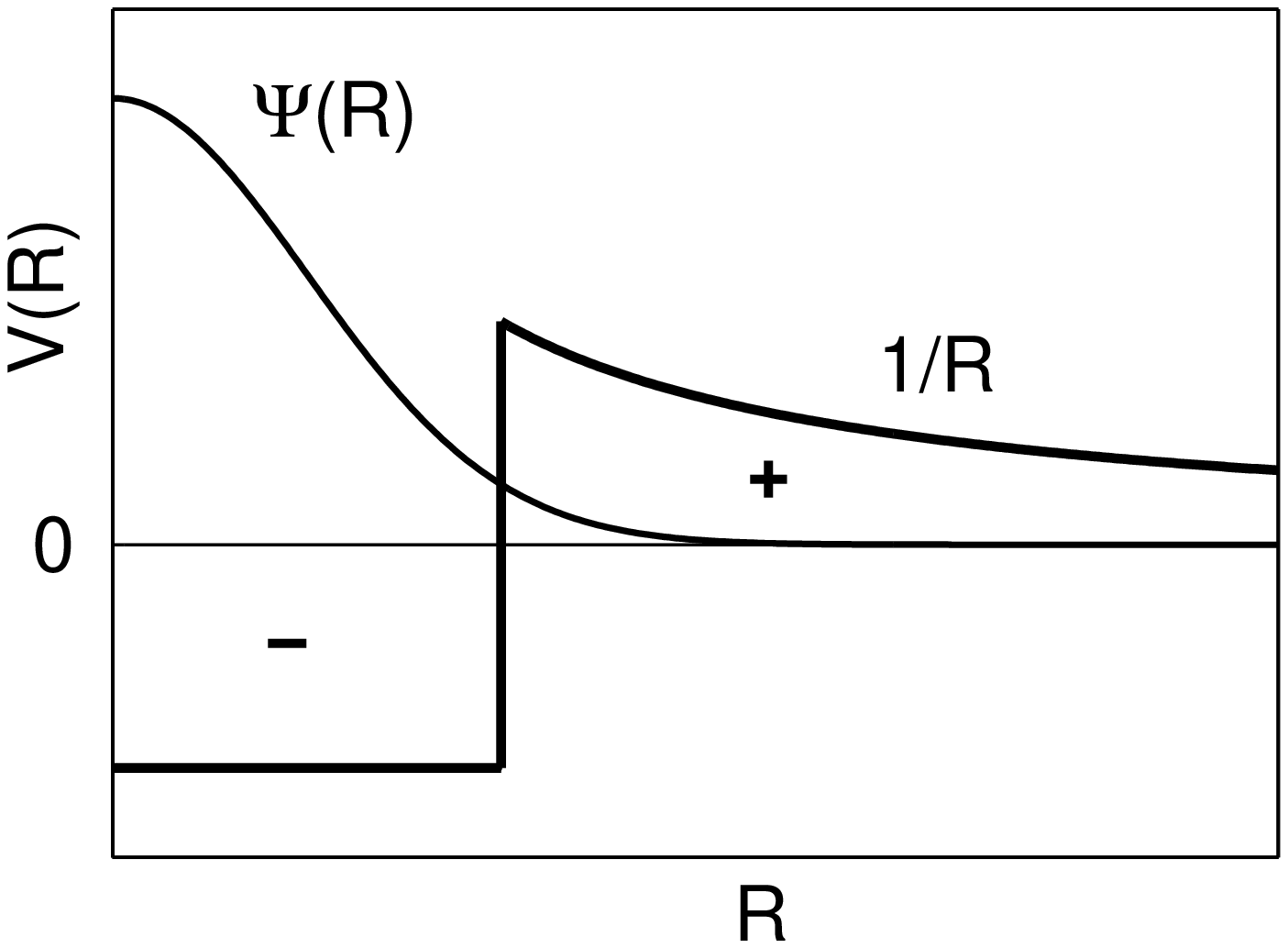}
 \includegraphics[width=6cm]{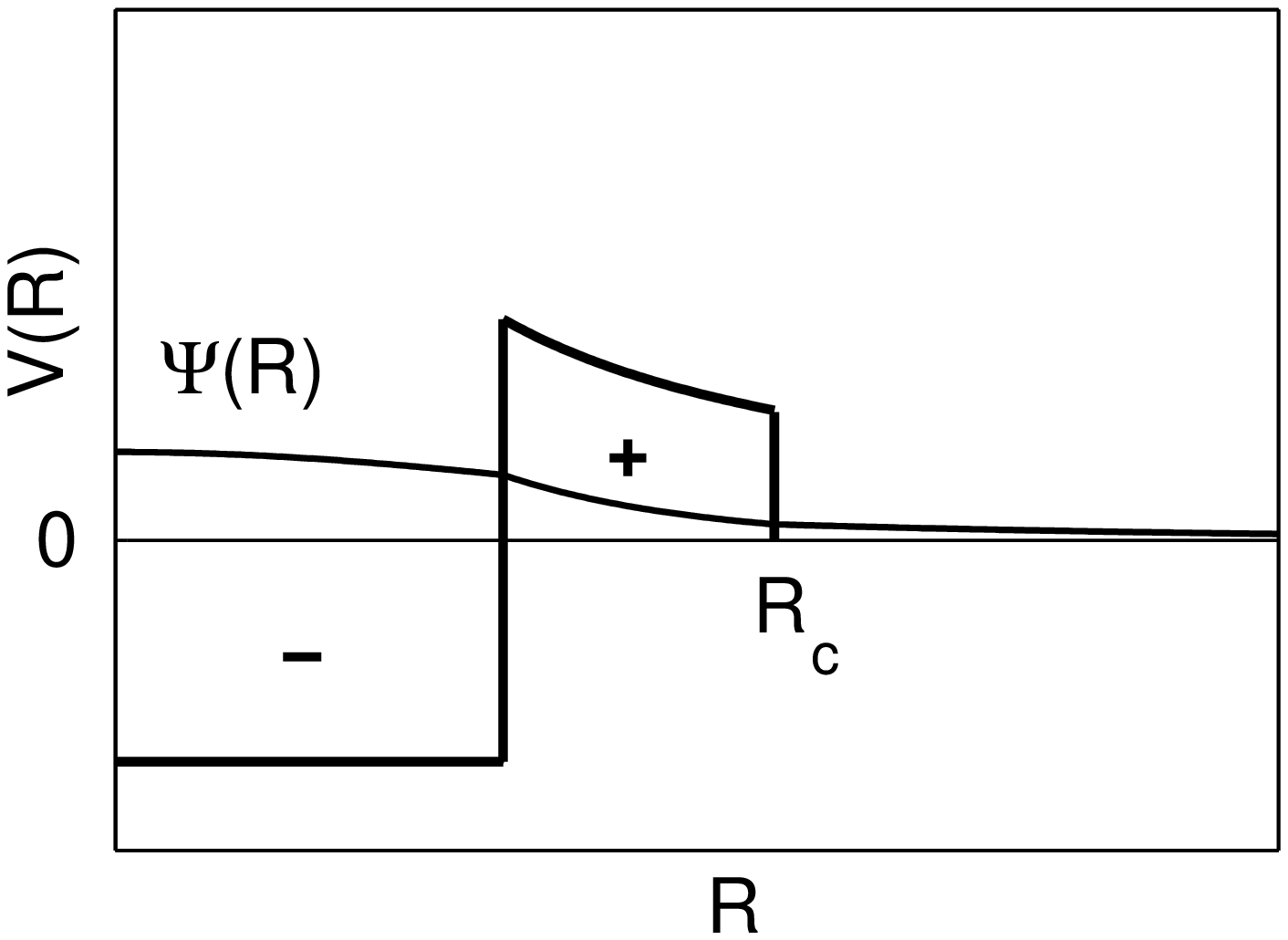}
 \caption{Sketched
behavior of the ground state wave functions approaching the
threshold in two potentials. Left: the potential has a positive
Coulomb tail. Right: the same potential is cut off at some distance
$R_c$. }
 \end{figure}


\section{Bound States near Threshold}\label{sec2}

In nature we can make the bound state of the system approach the
threshold by changing the number and the type of particles. In
theory we reproduce this behavior changing continuously some
parameters in the system. For example, in the case of ions
diminishing the atomic charge $Z$ down to the critical value $Z =
Z_{cr}$ makes the ground state approach the threshold
\cite{hogreve}. Here, for parameter, changing which we force the
states approach the threshold, we take the coupling constant of the
interaction.

In our analysis we shall consider the Hamiltonian of two particles
$H = H_0 + \lambda W$, where $H_0 = p^2$ is the free Hamiltonian (we
use the units where $\hbar = 1$ and $m = 1/2$), $W$ is the
interaction and $\lambda$ is a coupling constant. By decreasing
$\lambda$ we can lift any bound state to the continuum. For
convenience we shall consider only $W \in L^2 (\mathbb{R}^3 ) +
L^{\infty} (\mathbb{R}^3 )$ \cite{reed}. This is a large class of
interactions, which allows singularities not worse than
$r^{-\alpha}$ for $\alpha < 3/2$. In this case $H$ is self-adjoint
on the domain $D(H_0 ) = D(-\Delta )$ \cite{reed}. However, one
could extend our results to potentials having singularities of the
type $r^{-\alpha}$ for $\alpha < 2$.

Let us assume that there is a bound state having the energy
$E(\lambda) < 0$ for some value of the coupling constant $\lambda$ .
$E(\lambda)$ increases monotonously when $\lambda$ decreases and
eventually $E(\lambda)$ becomes zero for $\lambda = \lambda_{cr}$,
where $\lambda_{cr}$ is called the critical coupling constant
\cite{klaus}. In the following we show, using a simple example, how
the wave function spreads in the case of exponentially decaying
potentials. In this simple case the analyticity of the energy as a
function of $\lambda$ helps us to establish an upper bound on the
wave function
\begin{theorem}\label{th-1}
If there exist $A, a > 0 $ such that $|W| \leq A e^{-a|x|}$ and at
$\lambda = \lambda_{cr}$ there is no zero-energy bound state then
the following upper bound holds for the bound state $\psi $ having
the energy $E (\lambda)$ in the neighborhood of $E (\lambda_{cr} ) =
0$
\begin{equation}\label{psibound1}
    |\psi | \leq \frac{C|E|^{1/4} e^{-\sqrt{|E|}r}}{r},
\end{equation}
where $C > 0$ is some constant.
\end{theorem}
\begin{proof}  $\psi $ satisfies the integral equation $\psi = -[H - E]^{-1} W
\psi$, which could be rewritten as
\begin{equation}\label{lipsch}
\psi (x)= -\lambda \int dy \: \frac{ e^{ -\sqrt{|E|}|x-y| } W(y)
\psi(y) }{4\pi |x-y|}
\end{equation}
Using $|W| \leq A e^{-a|x|}$ and applying the Schwarz inequality to Eq.~(\ref{lipsch}) gives us
\begin{equation}\label{r2d2}
|\psi | \leq  \lambda \left< \psi | |W| | \psi\right>^{1/2} \left[
\int dy \: \frac{ A e^{-a |y| } e^{ -2\sqrt{|E|}|x-y|}  }{ |x-y|^2 }
\right]^{1/2} \leq \lambda \left< \psi | |W| | \psi\right>^{1/2}
\frac{C' e^{-\sqrt{|E|}r}}{r}
\end{equation}
where $C'$ is some constant and $r = |x|$. On the other hand, recall
\cite{klaus} that at $\lambda = \lambda_{cr} $ the energy
$E(\lambda)$ is analytic and can be expanded into convergent power
series $E(\lambda ) = \sum_{k=2}^\infty a_k (\lambda - \lambda_{cr}
)^k $, where $a_2 < 0$ ($a_1 = 0$ because by condition there is no
zero-energy bound state at $\lambda = \lambda_{cr} $). Applying the
Hellman-Feynman theorem $\left< \psi | H_0 | \psi \right> =
E(\lambda) - \lambda dE/d\lambda $ gives us $\left< \psi | H_0 |
\psi\right>  / |E|^{1/2} = O(1) $. Because $|W| \leq A e^{-a|x|}$
there must exist such constant $L$ that $|W| \leq L (2r)^{-2}$ and
thus by the uncertainty principle \cite{reed}, \cite{courant}
$\left< \phi | |W| | \phi \right> \leq L \left< \phi | H_0 | \phi
\right> $ for any $\phi \in D(H_0 )$. Thus $\left< \psi | |W| |
\psi\right> / |E|^{1/2} = O(1) $, which together with
Eq.~(\ref{r2d2}) proves the statement.
\end{proof}

As one can easily see, the function on the right-hand side of
Eq.~(\ref{psibound1}) dominates the wave function and spreads as $E
\to 0$ maintaining the constant norm independent of $E$. The
probability to find the particle in some fixed region of space goes
to zero.

Now, let us consider potentials with positive tails. Throughout the
paper we shall assume for such potentials that $W(x) \geq 0$ for
$|x| \geq R_0$, {\it i.e.}, that  outside some sphere the positive
part dominates. Below we present a simple criterion, which tells us
when bound states become bound states at the threshold. From the
discussion above and from Theorem~\ref{th-1} it is clear that the
strategy could be proving that as $\lambda \searrow \lambda_{cr}$
bound states remain confined in some region of space, {\it i.e.} do
not spread. One way to achieve this is to show that bound states are
dominated by some fixed function $g \in L^2$.

\begin{theorem}\label{th1}
Let $\lambda_n$ be a sequence of coupling constants and $\lambda_n
\searrow \lambda_{cr}$. The following is true (a) if for each
$\lambda_n$ there is a bound state $\phi_n$ with the energy $E_n$
such that $|\phi_n| \leq g$, where $g \in L^2$ and $E_n \to 0$ then
there exists a normalized bound state at the threshold, that is
$\phi_0 \in D(H)$ and $H(\lambda_{cr} ) \phi_0 = 0$. (b) if for each
$\lambda_n$ there are $m$ orthogonal bound states $\phi^{(m)}_n$
with the energies $E^{(m)}_n$ such that $|\phi^{(m)}_n | \leq g$,
where $g \in L^2$ and $\lim_{n\to \infty}E^{(m)}_n = 0$ then there
exist $m$ orthonormal bound states at the threshold.
\end{theorem}

The proof of this theorem, which follows the method in \cite{simon},
is given in the Appendix. Intuitively, this proposition is obvious.
If the states do not spread they should finally form some bound
state at the threshold. A similar theorem concerning many-body
systems appeared in Zhislin and Zhizhenkova \cite{zhislin}. There
the authors proved that if a minimizing sequence for the energy
functionals does not spread, then there exists a minimizer in $L^2$.
Their result could be explained from practitioners point of view.
Imagine that the system has no bound states with negative energy.
Then, if the function minimizing the energy functional, does not go
to zero as the number of basis functions increases, then there must
exist a zero-energy bound state.

Now let us see how the criterion in Theorem~2 works. First, we
separate positive and negative parts of the potential $W = W_+ - W_-
$, where $W_+ = \max (0, W)$ and $W_- = \max (0, -W)$ and $W_{\pm}
\geq 0$. The equation for the bound states reads $H(\lambda ) \phi =
- k^{2} \phi$, where $k \to 0$ as $\lambda \to \lambda_{cr}$. This
can be rewritten as
\begin{equation}\label{bs1}
    (H_0 + k^{2} + \lambda W_+ ) \phi = \lambda W_- \phi
\end{equation}
or equivalently
\begin{equation}\label{bs2}
    \phi = \lambda (H_0 + k^{2} + \lambda W_+ )^{-1} W_- \phi
\end{equation}

The operator $(H_0 + k^{2} + \lambda W_+ )^{-1}$ is an integral
operator, whose kernel is positive and real \cite{mark}. Thus we can
rewrite Eq.~(\ref{bs2}) as
\begin{equation}\label{pre1}
    | \phi | \leq 2 \lambda_{cr}  (H_0 + k^{2} + \lambda W_+ )^{-1} W_- | \phi | ,
\end{equation}
where because $\lambda \searrow \lambda_{cr}$ we have taken $\lambda
\leq 2 \lambda_{cr} $ without loss of generality. If we show that
the right-hand side of Eq.~(\ref{pre1}) is bounded by some fixed
square integrable function, then according to Theorem~\ref{th1} we
would have bound states at the threshold. The operator $(H_0 + k^{2}
+ \lambda W_+ )^{-1}$ is an integral operator, and its kernel is the
Green's function having two arguments. Because the function $W_- |
\phi |$ vanishes outside some sphere, the behavior of $| \phi |$ at
infinity is determined by the asymptotic of the Green's function
when the integration argument is fixed within the sphere. Thus to
find the asymptotic we need to derive upper and lower bounds on the
Green's function.

\section{Bounds on the Green's Functions} \label{sec3}

\subsection{Potential tails decaying as $1/r^2$ }

\subsubsection{Upper Bound}\label{usu1}

We introduce the function which would play the role of
potential's tail:
\begin{equation}\label{tail}
    \eta(A, R_0 ; x) = \left\{ \begin{array}{ll}
    0  & \quad \textrm{if $r < R_0 $} \\
    Ar^{-2} & \quad \textrm{if $ r \geq R_0$}, \\
    \end{array}
    \right.
\end{equation}

We are interested in the kernel of the integral operator $[H_0 + k^2
+ \eta (A, R_0 ; x)]^{-1}$ for $k$ real, which we denote as $G_k (A,
R_0 ; x,y)$. Note that the kernel of such operator is a positive
function,  continuous away from $x = y$ \cite{mark}. Our aim in this
section is to find the upper bound on $G_k (A,R_0 ; x,y)$. For that
we need the following lemma
\begin{lemma}\label{XX} Let $G_{1,2} (x,y)$ denote the integral
kernels of $[H_0 + k^2 + V_{1,2}]^{-1}$ and suppose $V_1 (x) \leq
V_2 (x)$. Then $G_2 (x,y ) \leq G_1 (x,y)$.
\end{lemma}
\begin{proof} Through the integral representation we get
\begin{equation}\label{eqXX}
G_2 (x,y) = G_1 (x,y) - \int d y' G_1 (x,y' ) [V_2 (y') - V_1 (y')]
G_2 (y' ,y)
\end{equation}
Eq.~(\ref{eqXX}) is the kernel representation of the equation
$(A+B)^{-1} = A^{-1} - A^{-1} B (A+ B)^{-1}$, where $A = H_0 + k^2 +
V_1$ and $B = V_2 - V_1$. Now, because $G_{1,2} (x,y)$ are positive
\cite{mark} and the potential difference $V_2 -V_1$ is non-negative,
the integral in Eq.~(\ref{eqXX}) must be non-negative and the
statement is proved. \end{proof}

There is another elucidating and more direct way to prove
Lemma~\ref{XX}. The proposition of the Lemma follows from the
Laplace transform and the Trotter product formula, which lie at the
heart of path-integrals \cite{reed}
\begin{gather}
(H_0 + k^2 + V_1 )^{-1} = \int^{\infty}_0 e^{-k^2 t} e^{-t(H_0 + V_1
)} dt \\
e^{-t(H_0 + V_1 )} = \slim_{m \to \infty} \left[ e^{-t H_0 /m }
e^{-t V_1  /m } \right]^m \label{garc}
\end{gather}
Because $e^{-tH_0}$ in Eq.~(\ref{garc}) has a positive kernel,
namely $(4 \pi t)^{-3/2} e^{-|x-y|^2 / 4t}$, the kernel of the
operator on the left in Eq.~(\ref{garc}) becomes smaller when $V_1$
is replaced by $V_2$.

The idea behind the upper bound on the Green's function is rather
simple. Suppose we have found such a function $F(A, R_0 , x)$,
independent of $k$, so that $G_k (A,R_0 ; x,0) \leq F(A,R_0 ; x) $
holds for all $k$ and $x$. First, we shall derive the upper bound in
terms of the function $F$, and then we shall determine $F$
explicitly.

Let us fix the functions $\tilde A (s), \tilde R_0 (s)$ so that the
following inequality holds
\begin{equation}\label{ceta}
    \eta (A , R_0 ; x  ) \geq \eta (\tilde A , \tilde R_0 ; x -s ),
\end{equation}
where $s$ is some fixed three--dimensional vector. From simple
geometric arguments it follows that Eq.~(\ref{ceta}) would be
satisfied if $\tilde A (s), \tilde R_0 (s)$ satisfy the inequalities
\begin{gather}
\tilde R_{0} (s) \geq R_0 + |s|   \label{tilR} \\ \tilde A (s) \leq
A \frac{\tilde R_{0}^2}{(\tilde R_{0} + |s|)^2} \label{tilA}
\end{gather}
Let us mention that the closer $\tilde A$ is to $A$ the better is
the asymptotic behavior of the bound, hence it is reasonable to take
$\tilde R_0$ large.

Translating the arguments one finds that $G_k (\tilde A, \tilde R_0
; x-s,y-s)$ is the integral kernel of the operator $[H_0 + k^2 +
\eta (\tilde A, \tilde R_0 ; x - s )]^{-1}$. Now, using
Eq.~(\ref{ceta}) and Lemma~\ref{XX} we obtain the upper bound
\begin{equation}\label{bgrf}
G_k (A, R_0 ; x,y) \leq G_k (\tilde A, \tilde R_0 ; x-s,y-s)
\end{equation}
Eq.~(\ref{bgrf}) is valid for all $s$, so we can put $s = y$, which
gives us
\begin{equation}\label{bgrf2}
G_k (A, R_0 ; x,y) \leq F (\tilde A (y), \tilde R_0 (y); x-y)
\end{equation}

It remains to find $F$, which is the upper bound on $G_k (A, R_0 ;
x,0)$. This is easy because $G_k (A, R_0 ; x,0)$ is spherically
symmetric in $x$. From now on for simplicity of notation we shall
drop $A, R_0$ in the arguments, writing, for example, $F(x)$ instead
of $F(A, R_0 ; x)$. First, we shall give a formal solution, then we
shall prove that this solution is indeed correct. By Lemma~\ref{XX}
$G_k (x , 0) \leq G_{k'} (x, 0)$ if $k' \leq k$, so we can take
$F(x) = \lim_{k \to 0} G_k (x , 0)$. Because $G_k (x , 0)$ is
continuous away from $x=0$ \cite{mark} and the functions $G_k (x ,
0)$ increase monotonically when $k \to 0$ the pointwise limit makes
sense. By Lemma~\ref{XX} $G_k (x , 0) \leq G^{(0)}_{k} (x, 0)$,
where $G^{(0)}_{k} (x, y) = (4\pi |x-y|)^{-1} \exp {(-k|x-y|)}$ is
the free propagator, {\it i.e.} the integral kernel of the operator
$[H_0 + k^2 ]^{-1}$. This means $F(x)$ is bounded away from $x=0$
and $F(x) \leq (4\pi r)^{-1}$. Because $G_k (x,y)$, formally
satisfies the equation $[H_0 + k^2 + \eta] G_k (x , y) =
\delta(x-y)$ one expects that $F(x)$ satisfies the equation
\begin{equation}\label{F}
    [ H_0 + \eta ]F = \delta (x)
\end{equation}
To find the solution of Eq.~(\ref{F}) we set
\begin{equation}\label{hatF}
    F = \frac 1{4\pi r} \times \left\{ \begin{array}{ll}
    1 + br  & \quad \textrm{if $r \leq R_0 $} \\
    cr^{-a} & \quad \textrm{if $ r \geq R_0$}, \\
    \end{array}
    \right.
\end{equation}
where $a$ is the positive root of the equation $a(a+1) = A$ and the
constants $b,c$ are fixed requiring, as usual, that $ F$ and its
derivative are continuous at $r = R_0$. This gives us
\begin{equation}\label{hatF2}
    F (A, R_0 ; x)= \frac 1{4\pi r} \times  \left\{ \begin{array}{ll}
    1 - R^{-1}_0 a (a+1)^{-1} r  & \quad \textrm{if $r \leq R_0 $} \\
    R_{0}^a (1+a)^{-1} r^{-a} & \quad \textrm{if $ r \geq R_0$} \\
    \end{array}
    \right.
\end{equation}
One can check that $F(x)$ defined by Eq.~(\ref{hatF2}) indeed
satisfies Eq.~(\ref{F}).

For completeness we give the accurate proof, which justifies
Eq.~(\ref{hatF2}).
\begin{lemma}\label{lemmap}
$F(x)$ defined as $F(x) = \lim_{k \to 0} G_k (x , 0)$ equals a.e.
the expression given by Eq.~(\ref{hatF2}).
\end{lemma}
\begin{proof}
The integral equation for the resolvent reads \cite{reed}
\begin{equation}\label{intres}
G_k (x, y) = G^{(0)}_k (x, y) - \int dy' G^{(0)}_k (x, y') \eta(y')
G_k (y', y)
\end{equation}
Substituting the expression for $G^{(0)}_k (x, y)$ and setting $y=0$
we obtain
\begin{equation}\label{intres2}
G_k (x, 0) = \frac {e^{-kr}}{4\pi r} - \frac 1{4\pi} \int dy'
\frac{e^{-k|x-y'|}}{|x-y'|} \eta(y') G_k (y', 0)
\end{equation}
Applying $\lim_{k \to 0}$ to both sides of Eq.~(\ref{intres2}) gives
us the integral equation
\begin{equation}\label{intF}
F(x) = \frac 1{4\pi r} - \frac 1{4\pi} \int dy' \frac
{\eta(y')}{|x-y'|}
 F(y')
\end{equation}
By simple substitution and calculating the integrals one can check
that $F$ given by Eq.~(\ref{hatF2}) indeed solves the integral
equation Eq.~(\ref{intF}). It remains to prove that no other
solution exists. Suppose there are two solutions and denote their
difference $Z = F_1 - F_2$. Then $Z$ satisfies the integral equation
\begin{equation}\label{Z}
Z(x) = - \frac 1{4\pi} \int dy' \frac 1{|x-y'|} \eta(y') Z(y')
\end{equation}
We need to show that $Z =0$ a.e. Let $\tilde Z = \eta^{1/2} Z$, then
$\tilde Z \in L^2$ by the dominated convergence theorem and $\|
\tilde Z \| \neq 0$, because otherwise from Eq.~(\ref{Z}) it follows
$Z = 0$ and we are done. From Eq.~(\ref{Z}) we obtain
\begin{equation}\label{Z2}
\int  dx dy \frac {\eta^{1/2} (x) \tilde Z(x) \eta^{1/2} (y) \tilde
Z(y)}{|x-y|} = - \| \tilde Z \|^2  <0
\end{equation}
But Eq.~(\ref{Z2}) cannot hold because $|x-y|^{-1}$ is the kernel of
the strictly positive operator. This means Eq.~(\ref{Z}) holds only
if $Z = 0$.\end{proof}

Now let us formulate the bound in the form required in
Sec.~\ref{mres}.
\begin{corollary}\label{ap-corol} Let $A > 3/4 $, then there exist $C>0$, $\delta > 0$ and $R$
such that for $|y| \leq R_0$ and $|x| \geq R$ the inequality holds
$G_k (A, R_0 ; x,y) \leq C |x|^{-3/2 - \delta}$.
\end{corollary}
\begin{proof} For $|y| \leq
R_0 $ we can fix the values $\tilde R_0$  and $\tilde A$ independent
of $y$. Both inequalities Eq.~(\ref{tilR}),~(\ref{tilA}) would be
satisfied if $\tilde R_0 \geq 2 R_0 $ and $\tilde A \leq A \tilde
R_{0}^2 (\tilde R_{0} + R_0 )^{-2}$. When $\tilde R_0$ becomes large
$\tilde A$ gets close to $A$, so we can fix the values of $\tilde
R_0$ and $\tilde A$ to ensure that the following inequality holds
$3/4 < \tilde A < A$. If we set $R = \tilde R_0 + R_0$, then for
$|y| \leq R_0$ and $|x| \geq R$ we have $|x-y| \geq \tilde R_0$ and
from Eq.~(\ref{bgrf2}), (\ref{hatF2}) we get
\begin{equation}\label{coroleq}
G_k (A, R_0 ; x,y) \leq (4 \pi |x| )^{-1} \tilde R^{\tilde a}_0 (1 +
\tilde a)^{-1} |x-y|^{-\tilde a} \leq C |x|^{-1-\tilde a},
\end{equation}
where $\tilde a$ is the positive root of the equation $\tilde a
(\tilde a+1) = \tilde A$, $\tilde a > 1/2$.\end{proof}

\subsubsection{Lower Bound}\label{usu2}

Here we shall briefly discuss how the same method can be applied to
construction of lower bounds. We would need this in
Section~\ref{mres} where we show that the ground state of potentials
decaying faster than $(3/4)r^{-2}$ spreads near the point of
critical binding. For that we need the following type of potential
\begin{equation}\label{new2}
    \xi( A , R_0 ,; x) = \left\{ \begin{array}{ll}
    V_0  & \quad \textrm{if $r < R_0 $} \\
    Ar^{-2} & \quad \textrm{if $ r \geq R_0$}, \\
    \end{array}
    \right.
\end{equation}
and we need the upper bound for Green's function of the operator
$\Xi_k (A, R_0 )= [H_0 + \xi + k^2 ]^{-1}$, which has the integral
kernel $\Xi_k (A, R_0 ; x,y)$. We shall derive the lower bound in
terms of the function $f_k (A, R_0 ; r)= \Xi_k (A, R_0 ; x,0)$,
which falls off at infinity and solves the following equation
\begin{equation}\label{e3}
    \left[ H_0 + \xi + k^2 \right] f_k = \delta(x)
\end{equation}
$f_k$ depends only on $r = |x|$ (because the potential is
spherically symmetric) and is a continuous function away from $r=0$.
By definition of $f_k$ we have $0 <  f_k \leq 1/(4 \pi r)$. Setting
$f_k = (4\pi r)^{-1} \hat f_k $ from Eq.~(\ref{e3}) we obtain the
equation on $\hat f_k$
\begin{equation}\label{katia}
    -\hat f_k '' + \xi \hat f_k + k^2 \hat f_k = 0
\end{equation}
with the boundary conditions $\hat f_k (0) = 1$ and $\hat f_k
(\infty ) = 0$. The function $0< \hat f_k (r) \leq 1$ comes out as a
solution of a simple radial equation and thus can be easily
calculated. As usual, one calculates the solutions $\hat f_k (r <
R_0 )$ and $\hat f_k (r > R_0 )$ and determines the constants so
that $\hat f_k $ and its derivative are continuous at $R_0$. The
following Lemma is useful for the lower bound.
\begin{lemma}\label{lem3}
For $r\geq R_0$ there exists $C_0$ independent of $k$ such that
\begin{equation}\label{katia6}
    \hat f_k (r  ) \geq C_0 e^{-kr} r^{-a}
\end{equation}
\end{lemma}

\begin{proof}
According to Eq.~(\ref{katia}) on the interval $[R_0 , \infty]$ the
function $\hat f_k $ satisfies the equation
\begin{equation}\label{katia2}
    -\hat f_k '' + Ar^{-2} \hat f_k + k^2 \hat f_k = 0
\end{equation}
Let us set $\hat f_k (r  ) = g_k (r)  e^{-kr} r^{-a}$. Then for $g_k
(r)$ on the interval $[R_0 , \infty]$ the equation becomes
\begin{equation}\label{katia3}
    -g_k '' + 2\left( ar^{-1} + k  \right) g_k ' = 2akr^{-1} g_k
\end{equation}
Because $f_k$ is positive $g_k$ should be also positive. Hence from
Eq.~(\ref{katia3}) we get
\begin{equation}\label{katia4}
g_k '' \leq 2\left( ar^{-1} + k  \right) g_k '
\end{equation}
We want to show that $g_k ' \geq 0$. Indeed, if on the contrary $g_k
' (y) < 0$ at some point $y$, then at this point due to
Eq.~(\ref{katia4}) $g_k '' (y) < 0$. Hence $g'$ is a monotonically
decreasing function for $r\geq y$. Thus from Eq.~(\ref{katia4}) we
conclude that $g_k '' \leq 2k g_k ' (y)$ for all $r > y$, {\em i.e.}
the second derivative is less than a fixed negative value, which
means that at some point $g_k $ becomes negative. Hence the
assumption was false and $g_k ' \geq 0$ holds. On the other hand,
$f_k > 0$ and as $k \to 0$ the function $f_k$ monotonically
increases in all points. Hence, there must exist $C_0
> 0$ such that $g_k (R_0 ) \geq C_0$. Together with $g_k ' \geq 0$ this
means that $g_k$ stays above $C_0 $ and Eq.~(\ref{katia6}) holds.
\end{proof}

Now we follow the above procedure and define $\tilde A , \tilde R_0$
as satisfying the inequality
\begin{equation}\label{new4}
\xi( \tilde A , \tilde R_0 ,; x-s) \geq \xi( A , R_0 ,; x)
\end{equation}
By geometrical arguments $\tilde A , \tilde R_0$ must satisfy
\begin{gather}
\tilde R_{0} (s) \geq R_0 + |s|   \label{new5} \\ \tilde A (s) \geq
A \frac{\tilde R_{0}^2}{(\tilde R_{0} + |s|)^2} \label{new55}
\end{gather}
Just as in the previous subsection through Eq.~(\ref{new4}) we
obtain the desired lower bound
\begin{equation}\label{new6}
\Xi_k (A, R_0 ; x,y) \geq \Xi_k (\tilde A, \tilde R_0 ; x-y , 0) =
f_k (\tilde A, \tilde R_0 ; |x-y| )
\end{equation}
%

Now suppose $A < 3/4 $. Looking at Eq.~(\ref{new5})--(\ref{new55})
one can see that we can fix $\tilde R_0$ and $\tilde A$ so that
$\tilde A < 3/4$ and Eq.~(\ref{new5})--(\ref{new55}) are valid. Then
from Eq.~(\ref{new6}) and Lemma~\ref{lem3} it is clear that there
exists a constant $C>0$ such that for $|x| \geq 2 \tilde R_0 $ and
for $|y| \leq R_0$ the following inequality holds
\begin{equation}\label{e34}
\Xi_k (A, R_0 ; x,y) \geq  f_k (\tilde A , \tilde R_0 ; |x-y|) \geq
C e^{-k|x|}|x|^{-3/2}
\end{equation}
We would need inequality Eq.~(\ref{e34}) in Sec.~\ref{mres}.

\subsection{Potential tails decaying as $1/r$ }\label{usu3}

Here we would like to apply the results of the previous section to
potentials with positive Coulomb tails. This helps to establish the
decay properties of eigenfunctions lying at the threshold. We shall
not present a detailed exposition, because everything is similar to
the previous section. One can follow in the steps of the previous
section and derive the bound in terms of the solution of the
equation $[H_0 + \eta'] F = \delta(r)$, where $\eta'$ is the Coulomb
tail. This however could not be expressed through elementary
functions, so we shall make a couple of simplifying approximations.
We shall consider the following potential tail.
\begin{equation}\label{tail'}
    \zeta(a, R_0 ; x) = \left\{ \begin{array}{ll}
    0  & \quad \textrm{if $r < R_0 $} \\
    (a^2 /4 )r^{-1} + (a/4) r^{-3/2} & \quad \textrm{if $ r \geq R_0$}, \\
    \end{array}
    \right.
\end{equation}

The repulsive Coulomb tail dominates in the potential of
Eq.~(\ref{tail'}) and one can choose the constants so that the
actual Coulomb tail is greater than the function in
Eq.~(\ref{tail'}). Let $G^{c}_k (a, R_0 ; x,y)$ be the integral
kernel of the operator $[H_0 + k^2 + \zeta (a, R_0 ; x)]^{-1}$,
where $c$ stands for Coulomb. The rest follows as above.

Let us fix the functions $\tilde a (s), \tilde R_0 (s)$ so that the
following inequality holds
\begin{equation}\label{zeta}
    \zeta (a , R_0 ; x  ) \geq \zeta (\tilde a , \tilde R_0 ; x -s ),
\end{equation}
Again from geometric arguments it follows that Eq.~(\ref{zeta})
would be satisfied if $\tilde a (s), \tilde R_0 (s)$ satisfy the
inequalities
\begin{gather}
\tilde R_{0} (s) \geq R_0 + |s|   \label{tilR'} \\
\tilde a (s) \leq a \left( \frac{\tilde R_{0}}{\tilde R_{0} + |s|}
\right)^{3/2} \label{tilA'}
\end{gather}
where the above conditions are obtained by directly applying
Eq.~(\ref{zeta}) to each positive term in the expression for $\zeta
(a , R_0 ; x  )$ given by Eq.~(\ref{tail'}).

Again let us define $F^{c} (x) = \lim_{k \to 0} G^{c}_k (x , 0)$,
which makes $F^{c} (x)$ satisfy the equation
\begin{equation}\label{F'}
    [ H_0 + \zeta ]F^{c} = \delta (x)
\end{equation}
As one can easily check, the solution of Eq.~(\ref{F'}) is given by
\begin{equation}\label{hatF2c}
    F^{c} (A, R_0 ; r)= \frac 1{4\pi r} \times \left\{ \begin{array}{ll}
    1 - \frac 1{R_0 + 2 \sqrt{R_0} / a } r  & \quad \textrm{if $r \leq R_0 $} \\
    \frac{e^{a \sqrt{R_0}}}{1 + (a/2)\sqrt{R_0}} e^{-a\sqrt{r}} & \quad \textrm{if $ r \geq R_0$} \\
    \end{array}
    \right.
\end{equation}

Finally, the upper bound reads
\begin{equation}\label{bgrf2'}
G^{c}_k (a, R_0 ;x,y) \leq F^{c} (\tilde a (y), \tilde R_0 (y); x-y)
,
\end{equation}
where $\tilde a $ and $\tilde R_0$  satisfy
Eq.~(\ref{tilR'})-(\ref{tilA'}). As in Corollary~\ref{ap-corol} from
Eq.~(\ref{bgrf2'}) we find that there exists such $R
> 0$ and $C> 0$ that
\begin{equation}\label{garbo}
G^{c}_k (a, R_0 ;x,y) \leq C e^{-a \sqrt{|x|}} \quad \textrm{if $|y|
\leq R_0 $ and $|x| \geq R$}
\end{equation}
As we have mentioned for potentials with positive tails the
asymptotic of the Green's function determines the fall-off behavior
of bound state wave functions. Hence the bound state wave functions
fall off at least as fast as $ e^{-a \sqrt{r}}$. Calculating in the
same way the lower bound one finds that this is the actual fall-off.

\section{Main Result}\label{mres}

Now we state the main result of this paper.
\begin{theorem}\label{thmain}
If there are such $R_0 $ and $A  > 3/4$ that $\lambda W_+ \geq
\eta(A, R_0 ; x)$ then at $\lambda = \lambda_{cr}$ all states that
hit the threshold at $\lambda = \lambda_{cr}$ become zero energy
bound states.
\end{theorem}
\begin{proof}
Let us define $G_{k} (A, R_0 ; x, y)$ the positive integral kernel
of the operator $[H_0 + k^2 + \eta ]^{-1}$. Then from
Eq.~(\ref{pre1}) and by Lemma~\ref{XX} we get the bound
\begin{equation}\label{punkt}
    | \phi | (x) \leq 2 \lambda_{cr}  \int_{|y| \leq R_0 } dy \: G_{k}
    (A, R_0 ; x, y ) W_- (y) | \phi | (y),
\end{equation}
where we have used $W_- (y)= 0$ for $|y| \leq R_0 $. Now we shall
use the upper bounds on the Green's function $G_{k} (A, R_0 ; x, y)$
derived in Se.~\ref{sec3}. For $|x| \geq R$ we can use
Corollary~\ref{ap-corol} to obtain from Eq.~(\ref{punkt})
\begin{equation}\label{g>}
 | \phi | (x) \leq 2 \lambda_{cr}  C |x|^{-3/2 -\delta} \int_{|y| \leq R_0 } dy
 \: W_- (y) | \phi | (y) \leq C_1 |x|^{-3/2 -\delta} \equiv g_{>} (x),
\end{equation}
where we have applied the Schwarz inequality and used $W \in L^2 +
L^{\infty}$. For $|x| \leq R$ we can use $G_{k} (A, R_0 ; x, y )
\leq (4\pi )^{-1 } |x-y|^{-1}$ to obtain from Eq.~(\ref{punkt})
\begin{gather}
 | \phi | (x) \leq 2 \lambda_{cr} (4\pi )^{-1 } \int_{|y| \leq R_0 } dy
 \: |x-y|^{-1} W_- (y) | \phi | (y) \nonumber \\
 \leq C_2  \left[ \int_{|y| \leq R_0 } dy
 \: |x-y|^{-2} W^{2}_- (y) \right]^{1/2}\equiv g_{<} (x) \label{g<},
\end{gather}
Thus we get $| \phi | (x) \leq g(x)$, where $g(x) = g_{<} (x)$ for
$|x| \leq R$ and $g(x) = g_{>} (x)$ for $|x| > R$. Because $g(x) \in
L^2 $ Theorem~\ref{th1} applies and the theorem is proved.
\end{proof}

One can construct simple examples, which show that the condition in
Theorem~\ref{thmain} is best possible. The following theorem is also
true.
\begin{theorem}\label{inv}
If there exists $R_0 $ such that
\begin{equation}
W(x) \leq  (3/4) |x|^{-2} \; \; \; {\rm for} \; \; |x| \geq R_0
\end{equation}
then the ground state cannot be bound at the threshold.
\end{theorem}
\begin{proof}
For simplicity we shall assume that there exists a constant $V_0$
such that $W_+ \leq V_0$. A proof by contradiction. Suppose that the
ground state $\psi_0 $ exists. Then the equation for the bound state
at the threshold can be written as $[H_0 + W_+ ]\psi_0 = W_-
\psi_0$. This is equivalent to the equation $[H_0 + W_+ + k^2
]\psi_0 = W_- \psi + k^2 \psi_0$, which in turn can be transformed
into the integral equation
\begin{equation}\label{new1}
    \psi_0 = [H_0 + W_+ + k^2 ]^{-1} W_- \psi_0 + k^2 [H_0 + W_+ + k^2 ]^{-1}
    \psi_0
\end{equation}
The ground state wave function is always nonnegative $\psi_0 \geq 0$
\cite{reed}. Because $W_- \geq 0$, $\psi_0 \geq 0$ in
Eq.~(\ref{new1}) on the right-hand side we have a sum of two
positive terms (the operator $[H_0 + W_+ + k^2 ]^{-1}$ has a
positive integral kernel). Hence we must have
\begin{equation}\label{vier}
\| [H_0+ W_+ + k^2 ]^{-1} W_- \psi_0 \| \leq 1
\end{equation}
for all $k$. Because the positive part of $W$ is bounded we have
$W_+ \leq \xi (A, R_0 ; x )$, where $\xi $ is defined by
Eq.~(\ref{new2}).

From Eq.~(\ref{vier}) and using Lemma~\ref{XX} we conclude $\| \Xi_k
(A, R_0 ) W_- \psi_0 \| \leq 1$ for all $k$, where $\Xi_k (A, R_0 )
= [H_0 + \xi + k^2 ]^{-1}$. Our aim is to prove $\lim_{k \to 0} \|
\Xi_k (A, R_0 ) W_- \psi_0 \| = \infty $ thus obtaining the desired
contradiction. We shall use the lower bound on $ \Xi_k (A, R_0 ;
x,y)$ from Sec.~\ref{usu2}. Let us fix $\tilde R_0$ as in the last
part of Sec.~\ref{usu2}. Then using the bound Eq.~(\ref{e34}) we
obtain for the square of the norm
\begin{gather}
\|\Xi_k W_- \psi_0 \|^2 \geq \int_{|y_1 | \leq R_0} \int_{|y_2 |
\leq R_0} dy_1 dy_2 \: W_- (y_1 ) \psi_0 (y_1 ) W_- (y_2 ) \psi_0
(y_2 ) \\
\times \int_{|x|  \geq 2\tilde R_0} dx \Xi_k (A, R_0 ; x, y_1 )\:
\Xi_k (A, R_0 ; x, y_2 ) \\
\geq M^2  C^2 \int_{|x|  \geq 2\tilde R_0} dx  |x|^{-3} e^{-2k|x|}
\label{eq1}
\end{gather}
where $M = \int_{|y| \leq R_0} dy W_- (y) \psi_0 (y) $ is some fixed
constant. Note that $M \neq 0$ because that would mean $ \psi_0 =0$.
It is clear that the right-hand side in Eq.~(\ref{eq1}) becomes
infinitely large as $k \to 0$ and thus $\| \Xi_k (A, R_0 ) W_-
\psi_0 \| \leq 1$ cannot hold for small $k$.
\end{proof}

 \begin{figure}[tb]
\includegraphics[width=6cm]{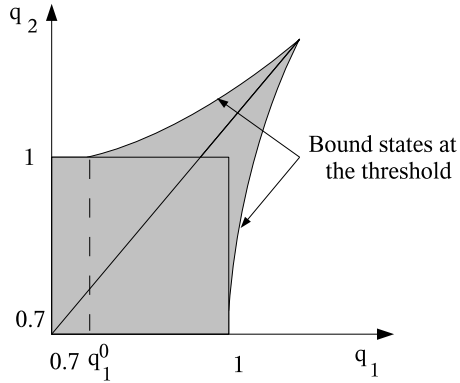}
 \caption{Typical
stability diagram (sketch) for three Coulomb charges $\{ -1, q_1 ,
q_2 \} $, the shaded area representing stable systems. On the arcs
of stability curve where either $q_1
> 1$ or $q_2
> 1$ there are bound states at the threshold.}
 \end{figure}

\section{Conclusions}

We have proposed the method to derive lower and upper bounds on the
Green's functions, which helps to determine the fall-off of bound
states. Using these bounds we have proved that potentials, whose
tails decay as $(2 \mu / \hbar^2 ) V > (3/4)r^{-2}$, where $\mu$ is
the reduced mass, absorb the eigenvalues, meaning that their bound
states do not spread and become bound states  at the threshold. We
have also found that ground states in potentials, whose tails decay
as $(2 \mu / \hbar^2 ) V < (3/4)r^{-2}$, always spread as they
approach the continuum.

These methods can be applied to the many-particle case, where it is
still not known, for which pair interactions between decaying
particles or clusters the bound state would become absorbed. The
difficulty is that it is hard to control the asymptotic of the
many-body bound state wave function. Using bounds on the Green's
functions derived here one can demonstrate \cite{later} that when
there is a long-range Coulomb repulsion between decaying components
(particles or clusters), then the bound state must get absorbed.

Two types of behavior, namely spreading and eigenvalue absorption
can be perfectly illustrated by a stability diagram for three
Coulomb charges, see \cite{martin}. When masses are fixed the
diagram has the form as in Fig.~2. If we consider, for example, the
upper arc, which is the stability border, it has the following
property \cite{martin}. Up to some non-zero value of $q^{0}_1$ the
stability border is given by the equation $q_2 = 1$. Then at the
point $\{ q^{0}_1, 1 \}$ the arc goes up. With the same method as
here it can be proved \cite{later} that if one approaches the
stability border from the side where $q_2 = 1$ then the ground state
spreads and there is no bound state at the threshold. On the
contrary, for the points on the arc the ground state becomes
absorbed, {\em i.e.} it does not spread and becomes a bound state at
the threshold. The reason is that on the stability border, where
$q_2 = 1$ the system decays into a neutral cluster and one charged
particle, and for $q_2 > 1$ both cluster and the particle are
charged positively. The resulting Coulomb repulsion between these
objects hinders the spreading of the wave function and the ground
state becomes absorbed. Note that it has already been proved
rigorously \cite{ostenhof}, that in the case of an infinite core and
two other masses being equal, the sharp point on the diagram, see
Fig.~2, has a bound state at the threshold. Our method helps to
extend this result to many particles.

\appendix

\section{Proof of Theorem~\ref{th1}}

\begin{proof}
Let us prove part (a). We follow the argument from \cite{simon}.
Because $\| \phi_n \| =1$ we can extract a weakly converging
subsequence (for which we reserve the same index $n$) such that
$\phi_n \stackrel{w}{\to} \phi_0$, where $\phi_0 \in L^2$. Because H
is self-adjoint in order to prove that $\phi_0 \in D(H)$ and $H
(\lambda_{cr} ) \phi_0 = 0$ it is enough to show that for every $f
\in D(H)$ we have $\bra H (\lambda_{cr}) f | \phi_0 \ket = 0$. The
latter we obtain as follows
\begin{eqnarray}
\bra H(\lambda_{cr} ) f | \phi_0 \ket = \lim_{n \rightarrow \infty}
\bra H (\lambda_{cr} ) f | \phi_n \ket =
\lim_{n \rightarrow \infty} [\bra H (\lambda_n ) f | \phi_n \ket -
(\lambda_n - \lambda_{cr} ) \bra Wf | \phi_n \ket ] =\\
\lim_{n \rightarrow \infty} \bra  f | H (\lambda_n ) \phi_n \ket =
\lim_{n \rightarrow \infty} E_n \bra f | \phi_n \ket = 0.
\end{eqnarray}
The only thing that remains to show is that $\phi_0 \neq 0$. We
shall prove this by contradiction assuming that $\phi_n
\stackrel{w}{\to} 0$. Let us introduce $\chi_R$, the characteristic
function of the interval $[0,R]$ ({\it i.e.} $\chi_R (x) = 1$ when
$|x| \in [0,R]$ and $\chi_R (x) = 0$ otherwise). Because $|\phi_n |
\leq g$ and $g \in L^2$ we can fix $R$ so that $ \bra \phi_n |
\chi_R | \phi_n \ket > 1/2$. We would like to show that for $\phi_n
\stackrel{w}{\to} 0$ the condition $ \bra \phi_n | \chi_R | \phi_n
\ket > 1/2$ cannot hold for large $n$. One way to do this is to use
that $\int |\nabla \phi_n |^2 dx \leq const$ and apply the
Rellich-Kondrashov lemma \cite{loss} giving $\chi_R \phi_n \to 0$
strongly. Or we can use the argument similar to the one in
\cite{simon}. Using the equation $(H_0 + 1)\phi_n = (E_n + 1)\phi_n
- \lambda_n W\phi_n$ we get $\phi_n = (E_n + 1) (H_0 + 1)^{-1}
\phi_n - \lambda_n (H_0 + 1)^{-1} W\phi_n$. Substituting this into $
\bra \phi_n | \chi_R | \phi_n \ket > 1/2 $ we obtain
\begin{equation}\label{nonsense}
(E_n + 1) \bra \phi_n | \chi_R (H_0 + 1)^{-1} \phi_n \ket -
\lambda_n \bra \phi_n | \chi_R (H_0 + 1)^{-1} W \phi_n \ket > 1/2
\end{equation}
The operators $\chi_R (H_0 + 1)^{-1}$ and $\chi_R (H_0 + 1)^{-1} W$
have square integrable kernels and are therefore compact. Acting on
weakly convergent sequences they make them converge strongly and
hence both terms on the left-hand side of Eq.~(\ref{nonsense}) go to
zero. Thus Eq.~(\ref{nonsense}) cannot hold for large $n$, which
proves (a).

Part (b) easily follows if we prove that from $\phi_n
\stackrel{w}{\to} \phi_0 $ follows $\phi_n \to \phi_0$ in norm.
Indeed, for each $\lambda_n$ there are bound states $\phi^{(i)}_n$
(i = 1, \ldots, m) satisfying $H(\lambda_n ) \phi^{(i)}_n =
E^{(i)}_n \phi^{(i)}_n$. Moreover $\bra \phi^{(i)}_n | \phi^{(k)}_n
\ket = \delta_{ik}$ and as $\lambda_n \searrow \lambda_{cr} $ the
energies go to zero $E^{(i)}_n \to 0$. In this case since
$|\phi^{(i)}_n | < g \in L^2$ there are $m$ bound states at the
threshold, $\phi^{(i)}_n \stackrel{w}{\to} \phi^{(i)}_0 $. Because
this convergence is in norm $\bra \phi^{(i)}_0 | \phi^{(k)}_0 \ket =
\delta_{ik}$ holds.

To prove that from $\phi_n \stackrel{w}{\to} \phi_0 $ follows
$\phi_n \to \phi_0$ in norm let us define $\xi_n = \phi_n - \phi_0$,
then $\xi_n \stackrel{w}{\to} 0 $ and we would like to show that $\|
\xi_n \| \to 0$. A proof by contradiction. If not then there must
exist a constant $a > 0$ and a subsequence (for which we again
reserve the same index $n$) such that $\| \xi_n \|^2 > a$. Again
because $| \phi_n | \leq g$ and $\phi_0 \in L^2 $ we can fix $R$ so
that $\bra \xi_n | \chi_R | \xi_n \ket > a/2$. We have $\xi_n \in
D(H)$ and $(H_0 + \lambda_n W )\xi_n = E_n \phi_n + (\lambda_{cr} -
\lambda_n ) W \phi_0$. From this equation we easily get
\begin{equation}\label{c1}
\xi_n = (H_0 + 1)^{-1} \xi_n - \lambda_n (H_0 + 1)^{-1} W \xi_n +
E_n (H_0 + 1)^{-1} \phi_n + (\lambda_{cr} - \lambda_n ) (H_0 +
1)^{-1} W \phi_0
\end{equation}
Substituting one $\xi_n $ from Eq.~(\ref{c1}) into $(\xi_n , \chi_R
\xi_n ) $ and using that $\chi_R (H_0 + 1)^{-1}$ and $\chi_R (H_0 +
1)^{-1} W$ are compact and $\xi_n \stackrel{w}{\to} 0 $ we obtain
$\bra \xi_n | \chi_R | \xi_n \ket \to 0$. This is a contradiction.
\end{proof}

\begin{acknowledgments} D.K. Gridnev expresses his gratitude to
H.~Hogreve for his interest to the problem and to the Humboldt
Fellowship for the financial support.
\end{acknowledgments}


\begin{thebibliography}{99}


\bibitem{hogreve}
H. Hogreve, J.~Phys.~B {\bf 31}, L439 (1998); Phys.~Scr.~{\bf 58},
25 (1998); private communication.

\bibitem{efimov}
T. Kraemer, {\em et.al.} Nature {\bf 440}, 315 (2006)

\bibitem{zhukov}
M.V. Zhukov, B.V. Danilin, D.V. Fedorov, J.M. Bang, I.J. Thompson
and J.S. Vaagen, Phys.~Rep. {\bf 231}, 151 (1993).

\bibitem{dimer}
F. Lou, C.F. Giese and W.R. Gentry, J.~Chem.~Phys. {\bf 104}, 1151
(1996)

\bibitem{fedorov}
D.~V.~Fedorov, A.~S.~Jensen, and K.~Riisager, Phys.~Rev.~C~{\bf 49},
201 (1994); A.~S.~Jensen, K.~Riisager, and D.~V.~Fedorov,
Rev.~Mod.~Phys.~{\bf 76} 215 (2004); K.~Riisager, D.~V.~Fedorov and
A.~S.~Jensen, Europhys.~Lett.~{\bf 49}, 547 (2000).

\bibitem{klaus}
M.~Klaus and B.~Simon, Ann.~Phys.~{\bf 130}, 251 (1980).


\bibitem{amer}
R.~K.~P. Zia, R.~Lipowski and D.M.~Kroll, Am.~J.~Phys.~{\bf 56}, 160
(1998).

\bibitem{gest}
D.~Bolle, F.~Gesztesy and W.Schweiger, J.~Math.~Phys~{\bf 26}, 1661
(1985).

\bibitem{newton}
R.~Newton, {\it Scattering Theory of Waves and Particles},
McGraw-Hill/New York 1966.


\bibitem{ostenhof}
M Hoffmann-Ostenhof, T Hoffmann-Ostenhof and B Simon,
J.~Phys.~A~{\bf 16}, 1125 (1983).

\bibitem{later}
D.~K.~Gridnev and M.~Garcia, to appear in Phys.~Rev.~A.

\bibitem{reed}
M. Reed and B. Simon, {\em Methods of Modern Mathematical Physics},
vol.~2-4, Academic Press/New York (1978)

\bibitem{courant}
R. Courant and D. Hilbert, {\em Methods of Mathematical Physics},
Interscience Publishers, New York, (1953), vol.~1, p. 446

\bibitem{simon}
B.~Simon, J.~Functional Analysis~{\bf 25}, 338 (1977).

\bibitem{zhislin}
G.~M.~Zhislin, Trudy~Mosk.~Mat.~Ob{\v s}{\v c}. {\bf 9}, 81 (1960);
E.~F.~Zhizhenkova and G.~M.~Zhislin, Trudy~Mosk.~Mat.~Ob{\v s}{\v
c}. {\bf 9}, 121 (1960)

\bibitem{loss}
E.~H.~Lieb and M.~Loss, {\it Analysis}, Amer.~Math.~Soc. second
edition, Providence, RI, 2001.


\bibitem{mark}
B.~Simon, Bull.~Amer.~Math.~Soc. {\bf 7}, 447 (1982)

\bibitem{martin}
A. Martin, J.M. Richard and T.T. Wu, Phys.~Rev.~{\bf A52}, 2557
(1995)

\end{thebibliography}
\end{document}